# Organometallic Hexahapto Functionalization of Single Layer Graphene as a Route to High Mobility Graphene Devices

Santanu Sarkar,† Hang Zhang,† Jhao-Wun Huang, Fenglin Wang, Elena Bekyarova, Chun Ning Lau,* and Robert C. Haddon*

[*] S. Sarkar, Dr. E. Bekyarova, Prof. R. C. Haddon
Center for Nanoscale Science and Engineering, Departments of Chemistry and Chemical & Environmental Engineering, University of California, Riverside, CA 92521-0403 (USA)
E-mail: haddon@ucr.edu

H. Zhang, J-W. Huang, F. Wang, Prof. C. N. Lau
Department of Physics and Astronomy, University of California, Riverside, CA 92521-0403 (USA).
E-mail: lau@physics.ucr.edu

†These authors contributed equally to the work.

**Abstract:** Organometallic hexahapto ($\eta^6$)-chromium metal complexation of single-layer graphene, which involves constructive rehybridization of the graphene $\pi$-system with the vacant chromium $d_\pi$ orbital, leads to field effect devices which retain a high degree of the mobility with enhanced on-off ratio. This $\eta^6$ mode of bonding is quite distinct from the modification in electronic structure induced by conventional covalent $\sigma$-bond formation with creation of $sp^3$ carbon centers in graphene lattice and this chemistry is reversible.

Pristine single layer graphene (SLG) has exceedingly high mobility, which is ~4,000–20,000 $cm^2/Vs$ for typical devices supported on $Si/SiO_2$ substrates, and may reach as high as 250,000 $cm^2/Vs$ in suspended devices at room temperature.[1] Such high mobilities make graphene an extremely attractive candidate for the next generation electronic materials. However, the absence of a band gap, which is necessary for digital electronics, presents a technological challenge. One effective approach to band gap engineering is the (partial) saturation of the valences of some of the conjugated carbon atoms.[2-17] Nitrophenyl functionalization, in which a fully rehybridized $sp^3$ carbon atom is created in the lattice, dramatically modifies the electronic and magnetic structure of graphene, with significantly reduced field effect mobility.[18-22] Since this type of functionalization scheme introduces resonant scatters[23] into the graphene lattice, we refer to this as destructive rehybridization.[24]

Most approaches for chemical modification of graphene involve the creation of $sp^3$ carbon centers at the cost of conjugated $sp^2$ carbon atoms in the graphene lattice. We have recently investigated the application of organometallic chemistry by studying the covalent hexahapto modification of graphitic surfaces with zero-valent transition metals such as chromium.[12, 25] The formation of the hexahapto ($\eta^6$)-arene–metal bond leads to very little structural reorganization of the $\pi$-system. In the reaction of the zero-valent chromium metal with graphene, the vacant $d_\pi$ orbital of the metal (chromium) constructively overlaps with the occupied $\pi$-orbitals of graphene, without removing any of the $sp^2$ carbon atoms from conjugation.[12, 25] Previously we have shown that the formation of such bis-hexahapto transition metal bonds between the conjugated surfaces of the benzenoid ring systems present in the surfaces of graphene and carbon nanotubes can dramatically change their electrical properties.[12, 24-27] These prior works focus on using the bis-hexahapto-metal bond as an interconnect for electrical transport between the conjugated surfaces, thereby increasing the dimensionality of the carbon nanotube and graphene materials and thus we were concerned with the use of the bis-hexahapto-metal bond as a conduit for electron transport between surfaces. In contrast, the goal of the present study is to investigate the effect of the hexahapto-bonded chromium atoms on the electronic properties of graphene itself (within the plane of a single layer), by using mono-hexahapto-metal bonds to the graphene surface.

Single layer graphene (SLG) flakes used in this study were extracted from bulk graphite using a standard mechanical exfoliation method and placed on a Si substrate with 300 nm $SiO_2$. Contacts consisting of 10 nm of Cr and 150 nm of Au were deposited on SLG by e-beam lithography. The devices were then annealed in vacuum by passing a high current for a short time to remove contaminants from the surface.[28] After characterization the devices are immersed in a chromium hexacarbonyl solution for organometallic functionalization.

Three different functionalization approaches (see Experimental Section for details) are employed to chemically modify the graphene flakes as shown in Figure 1. In the first method (method ***A***), SLG devices are functionalized in a solution of chromium hexacarbonyl [$Cr(CO)_6$] in dibutyl ether/tetrahydrofuran under refluxing conditions (140 $^{o}$C, 48 h). In the second method (method ***B***), the SLG devices were immersed in a solution of $Cr(CO)_6$ as in method ***A***, but in presence of an additional ligand, naphthalene (80 $^{o}$C for 12 h). The naphthalene was added in order to form the labile complex (naphthalene)$Cr(CO)_3$ complex (resulting from the haptotropic slippage of the naphthalene ligand from $\eta^6$- to $\eta^4$- or $\eta^2$-coordination) in-situ,[29-31] which is known to be a very effective reagent for the transfer of the –$Cr(CO)_3$ group between ligands. In our experience there is a facile arene exchange reaction between naphthalene and the more reactive graphene layer, which allows the reaction to proceed at relatively low temperature. In the third method (method ***C***), the SLG device is functionalized using a solution of tris(acetonitrile) tricarbonylchromium(0) [$Cr(CO)_3(CH_3CN)_3$][31-34] in THF (40 $^{o}$C, 6 h).

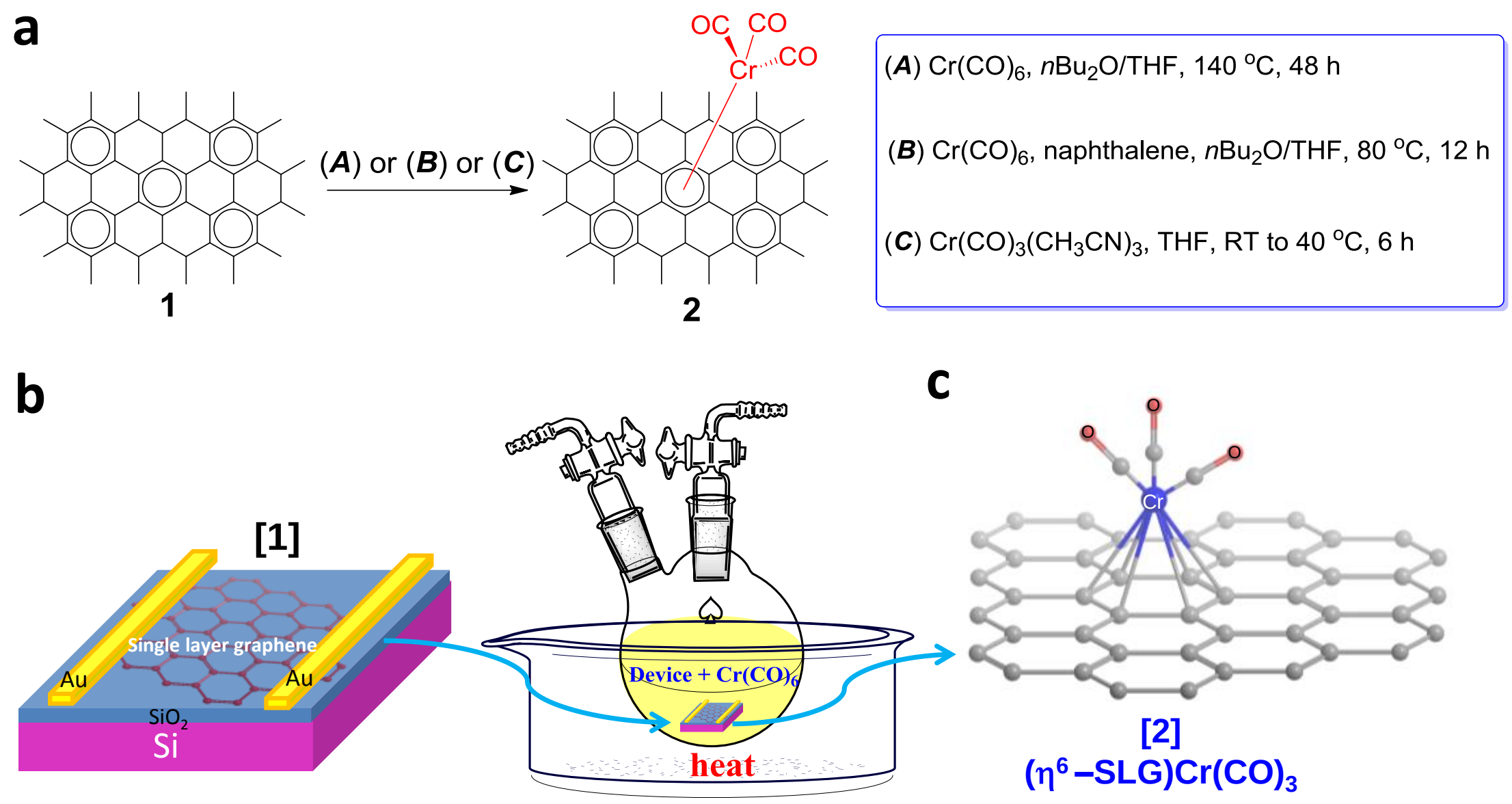

**Figure 1.** Organometallic functionalization of single-layer graphene devices (SLG, **1**): (a) Schematics of functionalization approaches using three different reaction routes to obtain hexahapto-chromium complex, ($\eta^6$-SLG)$Cr(CO)_3$ (**2**); routes: ***A***: $Cr(CO)_6$, *n*-$Bu_2O$/THF, 140 °C, 48 h, under argon, ***B***: $Cr(CO)_6$, naphthalene, *n*-$Bu_2O$/THF, 80 °C, 12 h, under argon, and ***C***: $Cr(CO)_3(CH_3CN)_3$, THF, room temperature to 40 °C, 6 h, under argon. (b) Illustration of the graphene device and the functionalization process; and (c) Three-dimensional model of the ($\eta^6$-SLG)$Cr(CO)_3$ organometallic complex.

To probe the effectiveness of the functionalization approaches the graphene sheets were characterized with Raman spectroscopy before and after the reaction ($\lambda_{ex}$ = 532 nm, Nicolet Almega XR). The Raman spectra of the pristine single layer graphene (SLG, Figure 2a-i) shows the characteristic G-band (1585 $cm^{-1}$) and 2D-band (2680 $cm^{-1}$), while organometallic covalent hexahapto ($\eta^6$-) functionalization leads to development of a D-band located at ~1345 $cm^{-1}$ with a relatively low intensity (Figure 2a, ii-iv). The Raman measurements showed that all three methods were effective in the formation of Cr-complexed graphene, although method ***C*** was found to provide functionalized graphene flakes with slightly weaker D-band intensity, which may be due to a lower degree of hexahapto ($\eta^6$-) complexation. We also observe that the reactivity of the flakes towards hexahapto organometallic functionalization reaction was dependent on the number of graphene layers; analysis of the integrated $I_D/I_G$ ratios indicated that single-layer graphene (SLG) was more reactive than few-layer graphene (FLG) and HOPG was least reactive.

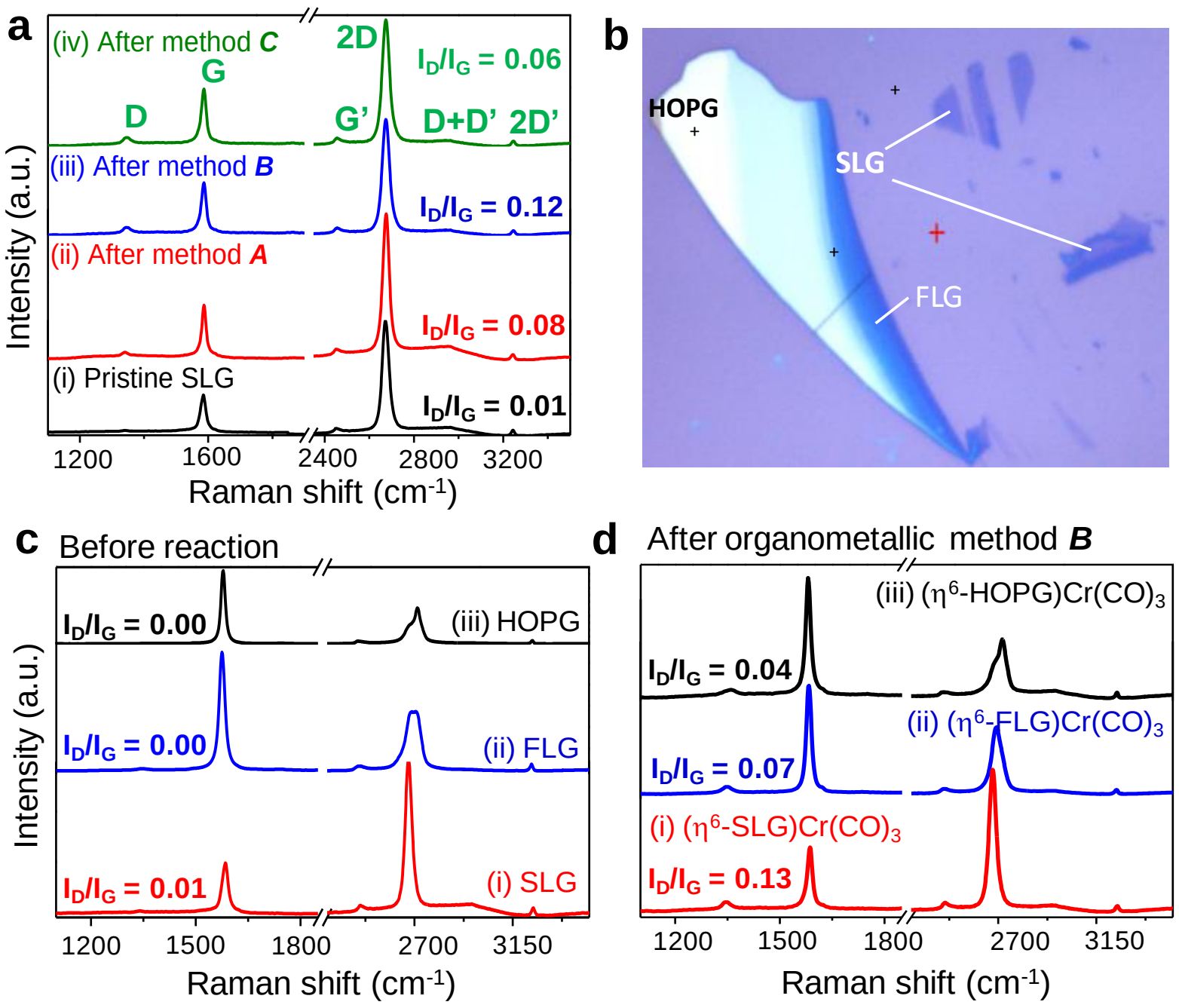

**Figure 2.** Organometallic functionalization of graphene and graphite. (a) Raman spectra of pristine SLG and chromium-functionalized graphene flakes, prepared by methods ***A***, ***B*** and ***C***. (b) Optical image of single-layer graphene (SLG), few layers graphene (FLG) and graphite (HOPG) on $SiO_2$/Si substrate. Contrast is enhanced by 30% for clarity. (c) Changes in chemical reactivity with stacking demonstrated by the evolution of Raman spectra before and after functionalization using method ***B*** on (i) SLG, (ii) FLG and (iii) HOPG.

In order to understand the effect of the hexahapto ($\eta^6$-) covalent binding of chromium atoms [–$Cr(CO)_3$ moieties] on the electronic properties of graphene, transport measurements were performed before and after functionalization of the devices. Because the Cr-graphene product can decompose at elevated temperatures,[12, 25] the ($\eta^6$-SLG)$Cr(CO)_3$ devices were characterized without annealing prior to the measurements. More than 10 devices were studied; in the present manuscript we report results on two devices prepared with Method ***A*** and Method ***C***, respectively. Based on the Raman spectra (Figure 2a) and the transport measurements, the former device appears to have a higher degree of functionalization. Figure 3a shows the zero-bias conductance ($G$) as a function of the applied gate voltage ($V_g$) of one pristine graphene device. The estimated room temperature field effect mobility of this device is $\mu$~4000 $cm^2$/Vs.

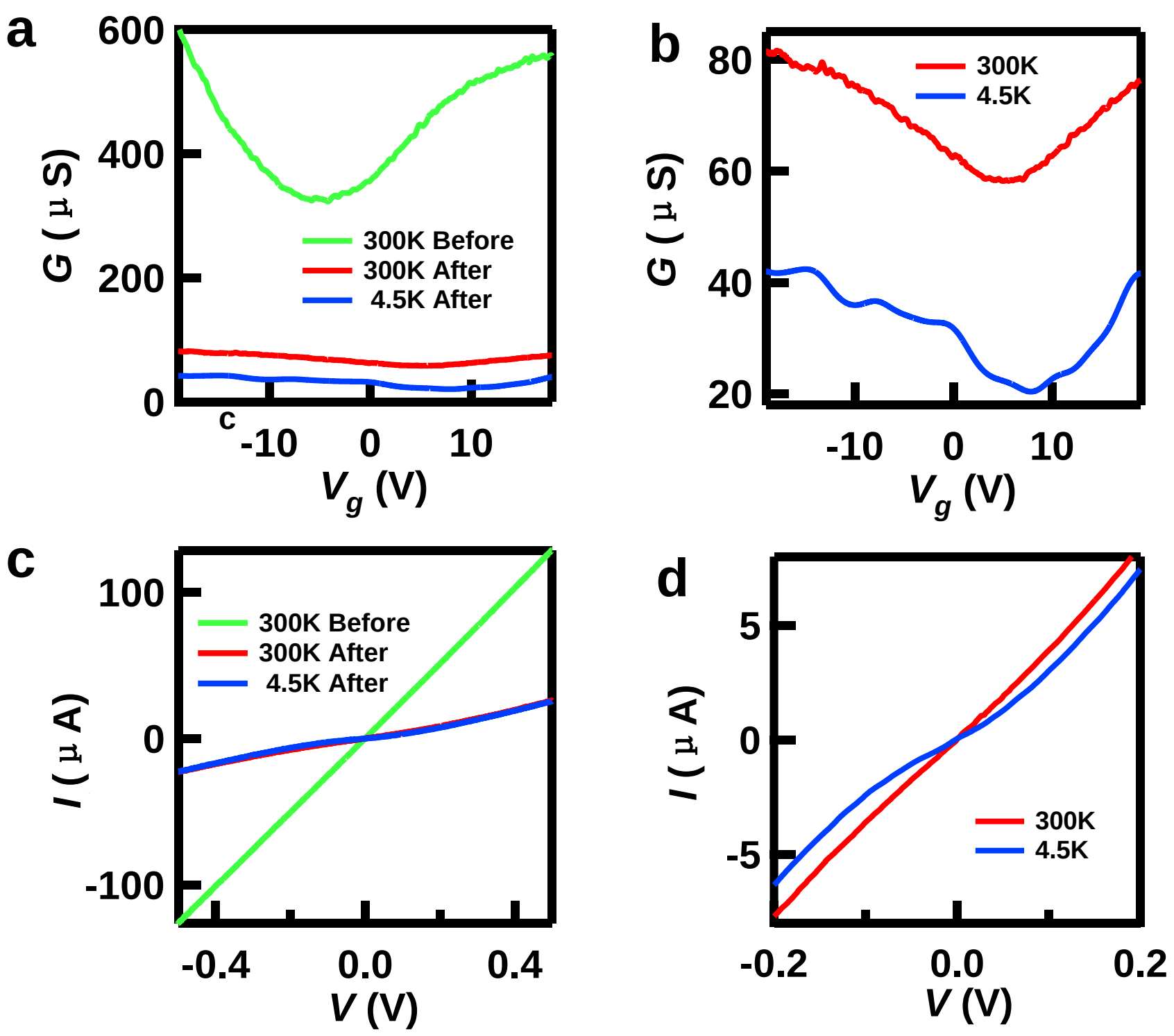

**Figure 3. (a)** *$G(V_g)$* characteristic from a device before and after functionalization. **(b)** *$G(V_g)$* characteristic of the functionalized device at 300 K and 4.5 K. **(c)** *I(V)* curves of the device before and after functionalization. **(d)** *I(V)* curves of the functionalized device at 300K and 4.5 K.

The pristine graphene device shows linear current-voltage (*I-V)* characteristics up to 0.5 V (Figure 3c). Both the *I-V* and *$G(V_g)$* curves of the pristine graphene device displayed weak temperature dependence, in agreement with previous experiments.[21] After functionalizing with chromium (method ***A***), the transport characteristics of the device changed significantly: the conductance of the device decreased 10 times (Figure 3a, 3b) and the *I(V)* curves became non-linear in the temperature interval 4 K – 300 K, as shown in Figure 3c and 3d. The estimated field effect mobility of the functionalized device is $\mu$~200 $cm^2$/Vs, while significantly diminished, is higher than previously reported values for functionalized graphene.[3, 21, 35, 36]

In order to explore the possibility of achieving high-mobility functionalized graphene devices, Cr-SLG devices derived by method ***C*** were characterized. Typically method ***C*** produced ($\eta^6$-SLG)$Cr(CO)_3$ devices with room temperature field effect mobility in the range of ~2,000 $cm^2/Vs$ and a current ON/OFF ratio of 5 to 13. The *G(V)* and *I(V)* curves at room temperature and 4.5 K of a weakly functionalized device [($\eta^6$-SLG)$Cr(CO)_3$ – method ***C***] are shown in Figure 4.

To understand the transport mechanism of the functionalized graphene device, the temperature dependence of conductance at the Dirac point and highly-doped regimes, were recorded in the range of 4 K to 300 K. The two most common transport mechanisms in functionalized devices are: (1) thermal activation over an energy gap ($2E_A$),[21] in which conductance decreases exponentially with the ratio between the activation energy $E_A$ and thermal energy $k_BT$, $G(T) = G_0 + A\exp(-E_A / k_BT)$ (equation 1), where $G_0$ = the constant background conductance, which is ascribed to the noise floor of the measurement setup, $k_B$ = Boltzmann constant, and (or) (2) variable range hopping (VRH), which displays a stretched exponential dependence $G(T) = A\exp\left[-(T_0 / T)^{\alpha}\right]$ (equation 2), where $T_0$ is a characteristic temperature and $\alpha$~ ½ to ¼ is the exponent.[21] To analyze the data, we plot *G* on a logarithmic scale as a function of $T^{-1}$ and $T^{-1/3}$[see Supporting Information (SI)]. Both plots exhibit some scatter but the thermally activated regression analysis (equation 1) gives values for the energy gap of $2E_A$ = 3 meV (Dirac point), $2E_A$ = 1 meV [highly doped regime (gate voltage of –42V)] (Figure S2, in SI); the largest energy gap that we observed in this study was for a device with a gap of $2E_A$ = 14 meV (Figure S3, in SI). Thus the data are consistent with the formation of true band gap of $2E_A \approx 10$ meV.[21] A possible complication in analyzing the transport data is the mobility (dynamic nature) of the chromium atoms [$-Cr(CO)_3$ moieties] on the graphene surface which may be evident in the data at high temperatures (Figure S2, in SI); such fluxional behavior has been observed in previous studies of polyaromatic hydrocarbon ligands,[38, 39] and this may be operative on the two-dimensional surface of the organometallic ($\eta^6$-SLG)$Cr(CO)_3$ complexes.[12]

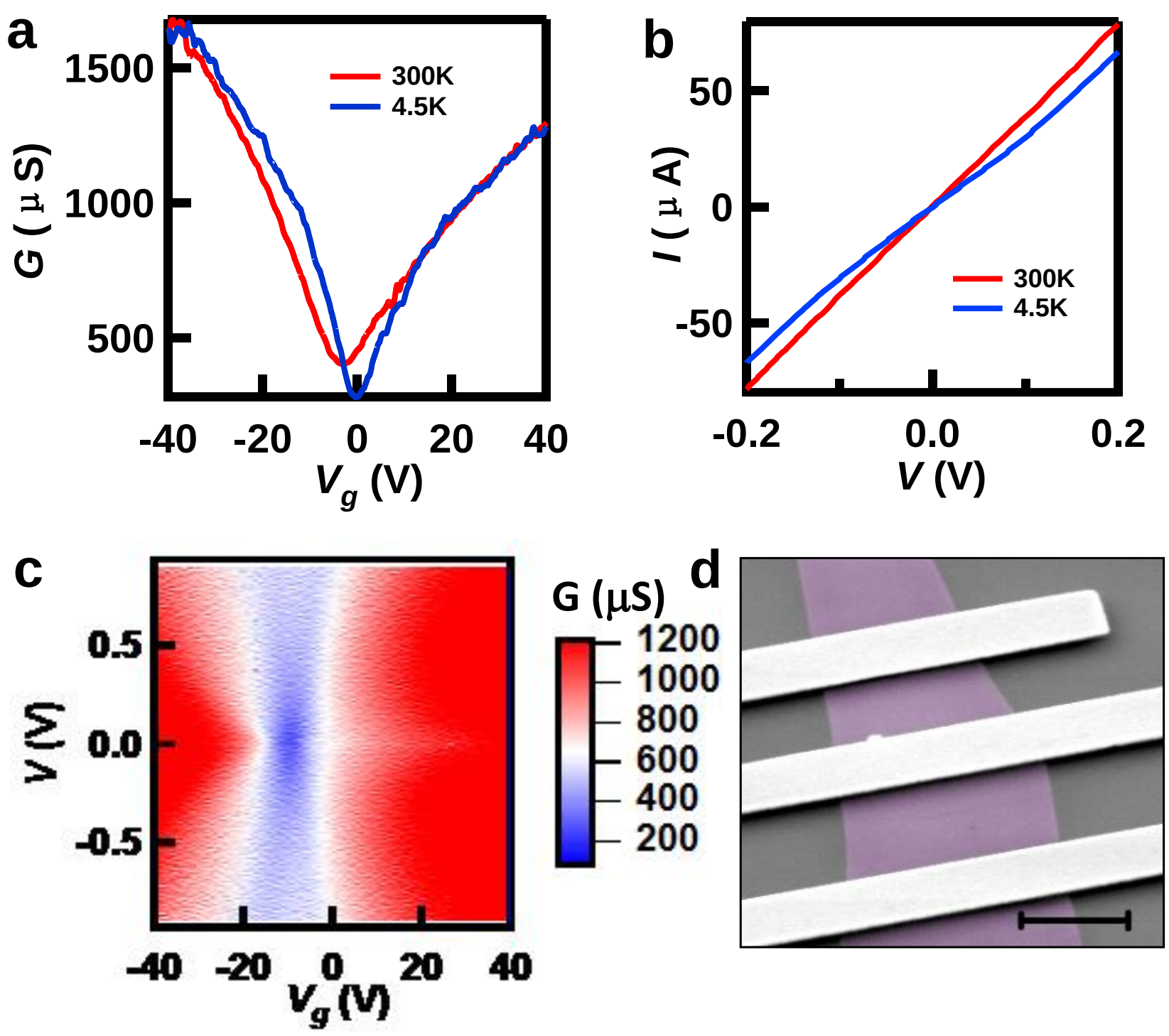

**Figure 4.** (a,b) *G(V)* characteristics and *I(V)* curves of a weakly functionalized device at 300 K and 4.5 K. The functionalized device has a mobility of ~2,000 $cm^2$/Vs at room temperature and ~3,500 $cm^2$/Vs at 4.5 K. (c) Conductance *G* as a function of bias *V* and gate $V_g$ at 4.5 K of the same device. (d) SEM image of a typical device (scale bar 2 μm).

We performed X-ray photoelectron spectroscopy (XPS) to estimate the coverage of the -$Cr(CO)_3$ units on the graphene surface. Because of the very small dimensions of the micromechanically exfoliated single-layer graphene (SLG) flakes and the fact that the presence of additional graphitic flakes on the silicon substrates is unavoidable, CVD-grown SLG (4 mm × 4 mm, on Cu-substrate) was prepared for the XPS experiments. The SLG samples were functionalized with chromium hexacarbonyl following the procedure described in Method ***C***. The survey spectrum of the functionalized samples in Figure 5 illustrates the doublet peak corresponding to Cr2p orbitals. The elemental composition was estimated from the areas of the peaks after Shirley background correction and the corresponding sensitivity factors. The analysis gave a C:Cr ratio of about 18 :1, which in the ideal case gives a structure such as that illustrated in the inset of Figure 5.

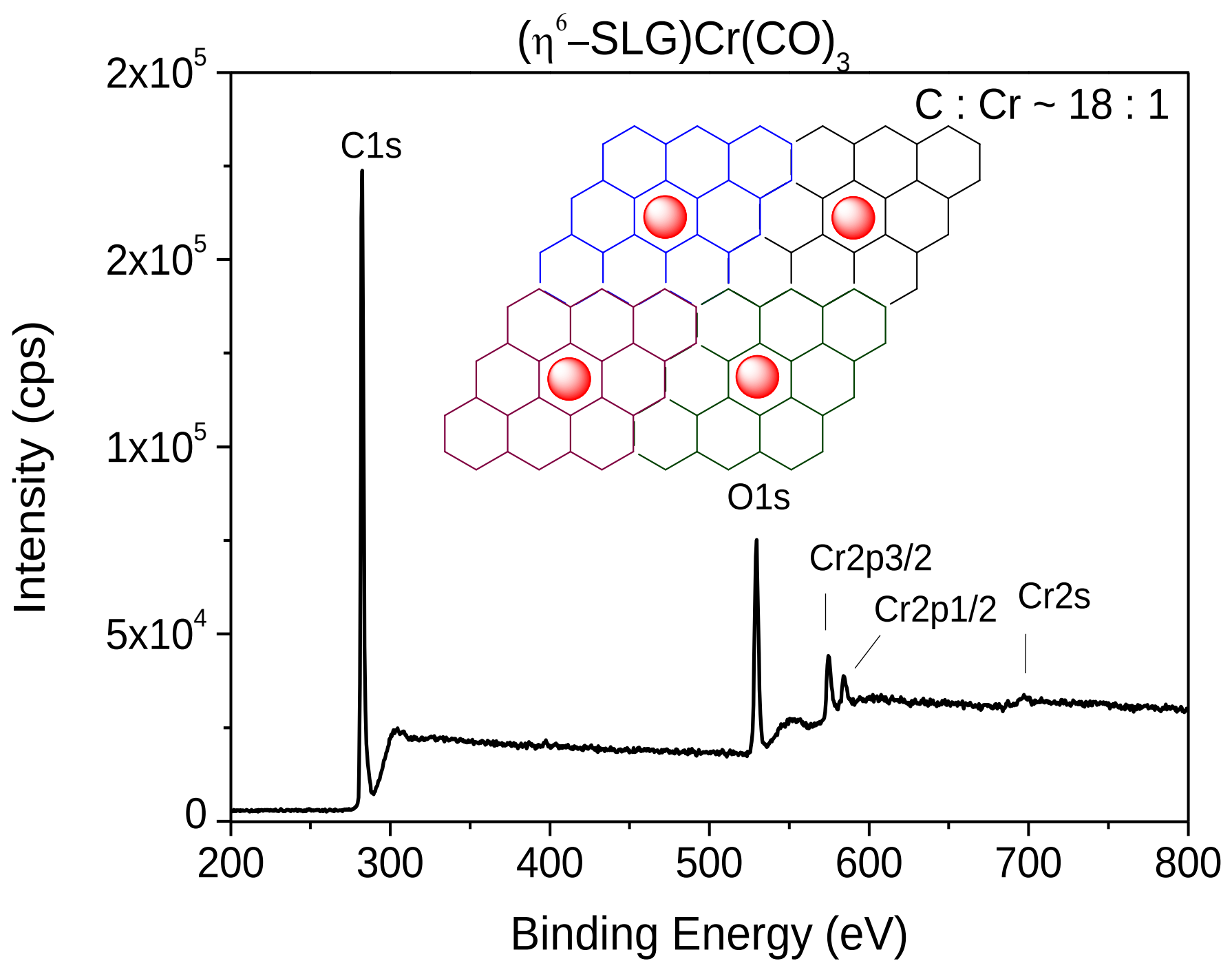

**Figure 5.** Survey spectrum of CVD-grown single-layer graphene (SLG) functionalized with chromium(0)tricarbonyl moieties. The inset shows the structure corresponding to the C:Cr ratio of 18:1 estimated from the C1s and Cr2p peaks, taking into account the sensitivity factors for carbon and chromium.

Another important characteristic of the chromium functionalization is its reversibility via decomplexation reactions. To achieve decomplexation the functionalized devices, ($\eta^6$-SLG)$Cr(CO)_3$ were exposed to an electron-rich ligand, such as anisole (Figure 6a). In a typical reaction, the device with an organometallic ($\eta^6$-SLG)$Cr(CO)_3$ complex was heated (150 °C) in presence of excess anisole (~10 mL) under argon for 12 hours. The device was then washed with chloroform, acetone, and hexane and dried under argon.

The complexation and decomplexation reactions were followed by Raman spectroscopy. As shown in Figure 6b, the intensity of the D-band was significantly reduced after the decomplexation reaction ($I_D/I_G = 0.03$). The small remaining D-band after the chemical reversal of the complexation reaction is presumably due to the generation of metal clusters or to the oxidation of the metal on the graphene lattice during heating in the organic solvent (tetrahydrofuran).

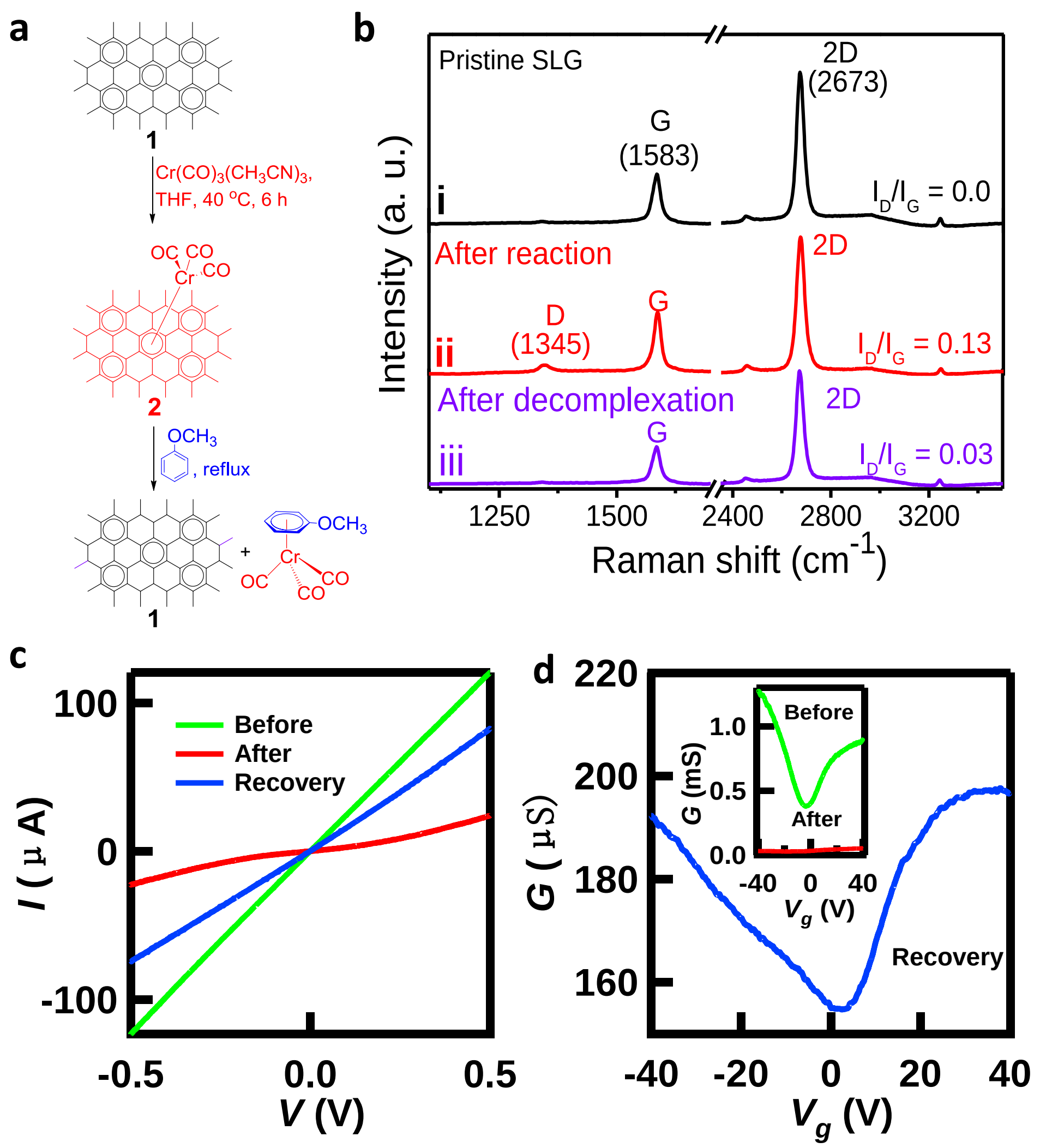

**Figure 6.** Decomplexation of chromium-graphene complexes. (a) Schematics of the complexation of the aromatic rings of graphene with –$Cr(CO)_3$ moieties by use of $Cr(CH_3CN)_3(CO)_3$ reagents (method ***C***), and decomplexation of the same using electron-rich ligand – anisole, to regenerate a clean graphene and ($\eta^6$-anisole)$Cr(CO)_3$. (b) Raman spectra of single layer graphene (SLG) – *i*: before reaction, *ii*: after reactions with $Cr(CO)_6$, naphthalene, *n*$Bu_2O$/THF, 80 °C, 12 h (method ***B***), and *iii*: after decomplexation. (c) *I(V)* curves and (d) *G* ($V_g$) curves of SLG device: green - pristine device, red – Cr-functionalized graphene device, blue – functionalized device after chemical recovery with anisole ligands. The measurements were performed at room temperature.

Mass spectroscopic (ESI-MS) analysis of the concentrated extract, which resulted from the competitive arene exchange reaction between ($\eta^6$-SLG)$Cr(CO)_3$ and anisole, led to the identification of a product corresponding to ($\eta^6$-anisole)$Cr(CO)_3$, which was detected with m/z = 243.9832. The transport measurements showed that the conductance and mobility of the devices were increased, although a complete recovery of the pristine device performance was not observed.

In summary, we find that the mono-hexahapto-chromium complexation of single layer graphene allows the fabrication of high performance chemically functionalized devices. We demonstrated that chemically modified graphene devices with a room-temperature field effect mobility in the range of $\mu$ ~200 - 2,000 $cm^2/Vs$ and an on/off ratio of 5 to 13 can be fabricated via $\eta^6$-metal complexation of graphene. Furthermore the graphene organometallic complexation chemistry may be reversed by treatment of the devices with electron rich ligands.  These graphene-metal complexes are potential candidates for advanced molecular wires,[40]  spintronics devices,[37] and organometallic catalyst supports.[12, 41] The finding that the in-plane transport properties are retained in the presence of mono-hexahapto-coordinated transition metals encourages the pursuit of this mode of bonding in 2-D and 3-D structures, which employ bis-hexahapto-metal complexation.[12, 24-27]

**Experimental Section**

*Materials and Characterization:* Chromium(0) hexacarbonyl [98%,  $Cr(CO)_6$, F.W. = 220.06, m.p. = 150 $^oC$, b.p. = 210 $^oC$ (decomposes)], tris(acetonitrile)tricarbonylchromium(0) [$Cr(CO)_3(CH_3CN)_3$, F.W. = 259.18, m.p. = 67-72 $^oC$ (decomposes)], naphthalene (F.W. = 128.17, m.p.= 80.26 $^oC$), *n*-dibutylether (b.p.= 140-142 $^oC$), anhydrous tetrahydrofuran (b.p.= 66 $^oC$), and anisole (b.p. = 154 $^oC$) were all obtained from Sigma-Aldrich. All chromium reagents have high vapor pressures and direct exposure to the reagents should therefore be avoided. Raman spectra were collected in a Nicolet Almega XR Dispersive Raman microscope with a laser excitation of 532 nm and with spectral resolution of 6 $cm^{-1}$ and spatial resolution of 0.7 $\mu$m. X-ray photoelectron spectroscopy (XPS) characterization was carried out by using a Kratos AXIS ULTRA$^{DLD}$ XPS system equipped with an Al K$\alpha$ monochromatic

X-ray source and a 165-mm electron energy hemispherical analyzer. Vacuum pressure was kept below $3 \times 10^{-9}$ torr during the acquisition. The survey spectra were recorded using 270 Watts of X-ray power, 80 pass energy, and 0.5 eV step size. A low-energy electron flood from a filament was used for charge neutralization. Scanning electron microscopy (SEM) images were acquired in a XL30-FEG SEM instrument. Electrospray ionization mass spectroscopy (ESI-MS) analysis was performed using the Agilent LCTOF instrument.

*Device Farbication:* Single-layer graphene flakes were isolated from bulk graphite by using standard micromechanical exfolaition technique and are placed on an oxidized silicon wafer (with 300 nm $SiO_2$). The contacts were deposited by e-beam lithography (Cr/Au – 10 nm/150 nm).

*Organometallic Complexation Reactions*: The hexahapto metal complexations reactions were perfomed at elevated temperatures under argon atmosphere using either chromium hexacarbonyls (in absence or presence of naphthalene as an additional ligand, method ***A*** and ***B*** respectively), or tris(acetonitrile)tricarbonylchromium (method ***C***).

***Method A:*** SLG devices were immersed in a chromium hexacarbonyl [$Cr(CO)_6$, 0.1M] solution in dibutyl ether/tetrahydrofuran (THF) (5:1) and refluxed under argon atmosphere at 140 $^{o}$C for 48 hours. After functionalization the graphene devices were washed carefully with THF.

***Method B:*** SLG devices were immersed in a solution of $Cr(CO)_6$ (as in method ***A***), with the addition of 0.25 equivalents of naphthalene ligand and heated to 80 $^{o}$C for 12 hours.

***Method C:*** SLG device was immersed in a solution of tris(acetonitrile) tricarbonylchromium(0) [$Cr(CO)_3(CH_3CN)_3$] in THF (~0.1 M) inside a glove-box and the reaction vessel was closed with rubber septum to maintain the argon atmosphere. The reaction vessel containing the graphene device and the solution were removed from the glove-box, connected to an argon line and heated slowly from room temperature to 40 $^{o}$C for 6 hours.

This procedure required rigorous exclusion of the atmosphere in order to avoid doping the graphene as a result of the decomposition of the chromium reagent to chromium oxide.

*Decomplexation Reactions*: In a typical reaction, the organometallic ($\eta^6$-SLG)$Cr(CO)_3$ complex was refluxed (at 150 $^oC$) in presence of excess anisole (~10 mL) under an atmosphere of argon overnight. The resulting SLG flake was washed with chloroform, acetone, and hexane; dried with gentle flow of argon. The decomplexation product of ($\eta^6$-SLG)$Cr(CO)_3$ with anisole yielded a light yellow solution, which was analyzed (after concentration with a rotary evaporator) using electrospray ionization mass spectroscopy (ESI-MS), which confirmed the chemical composition as ($\eta^6$-anisole)$Cr(CO)_3$ (chemical formula: $C_{10}H_8CrO_4$, F.W. = 243.98). ESI-MS data shows: m/z = 243.9832 [{(M+H)+[–H]}, diff(ppm): 3.88, $C_{10}H_8CrO_4$, calculated m/z = 243.9822], and m/z = 244.9904 $[(M+H)]^+$, diff(ppm): 1.36, $C_{10}H_9CrO_4$, calculated m/z = 244.9901].

*Transport Measurements of the Devices:* The devices were placed into a custom-built helium cryostat. All the measurements were performed in a high vacuum enviornment. The temperature of the devices was measured with a semiconductor thermometer mounted in close proximity to the chip carriers. Data were acquired by National Instrument PCI-6251 card controlled by a C++ based program.

**Acknowledgements**

This material is based on research sponsored by the Defense Microelectronics Activity (DMEA) under agreement number H94003-10-2-1003. XPS measurements were recorded at UCR with support from the National Science Foundation under Grant DMR-0958796.

# Supporting Information

## Organometallic Hexahapto Functionalization of Single Layer Graphene as a Route to High Mobility Graphene Devices

*By* Santanu Sarkar,[†] Hang Zhang,[†] Jhao-Wun Huang, Fenglin Wang, Elena Bekyarova, Chun Ning Lau,* *and* Robert C. Haddon*

**Temperature Dependence of Conductance:**

**Pristine Single Layer Graphene:** Typically, the conductance of pristine single-layer graphene devices has a very weak temperature dependence, as shown in Figure S1.

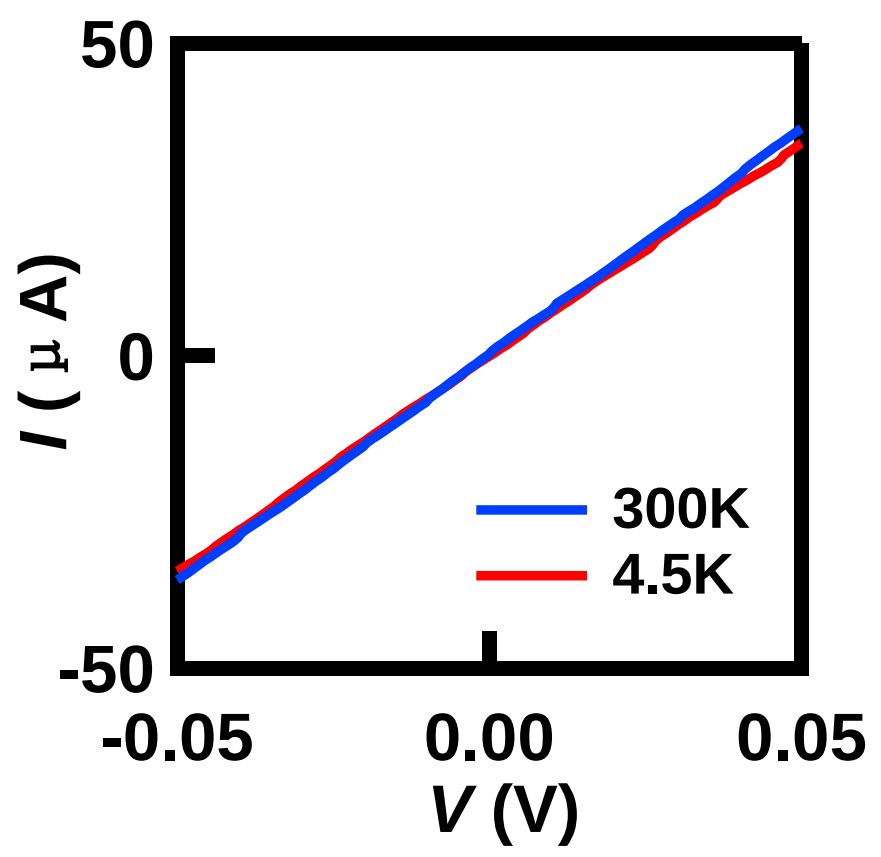

**Figure S1.** I-V curves at Dirac point of a pristine graphene device at T = 300 K and 4.5 K.

**Chromium Functionalized Single Layer Graphene:** A moderate temperature dependence of conductance is observed for the chromium functionalized single-layer graphene devices. The zero bias conductance of a Cr functionalized graphene device as a function of temperature in different doping regimes is illustrated in **Figure S2** and **S3**. The conductance data is plotted as a function of $T^{-1}$ and $T^{-1/3}$.

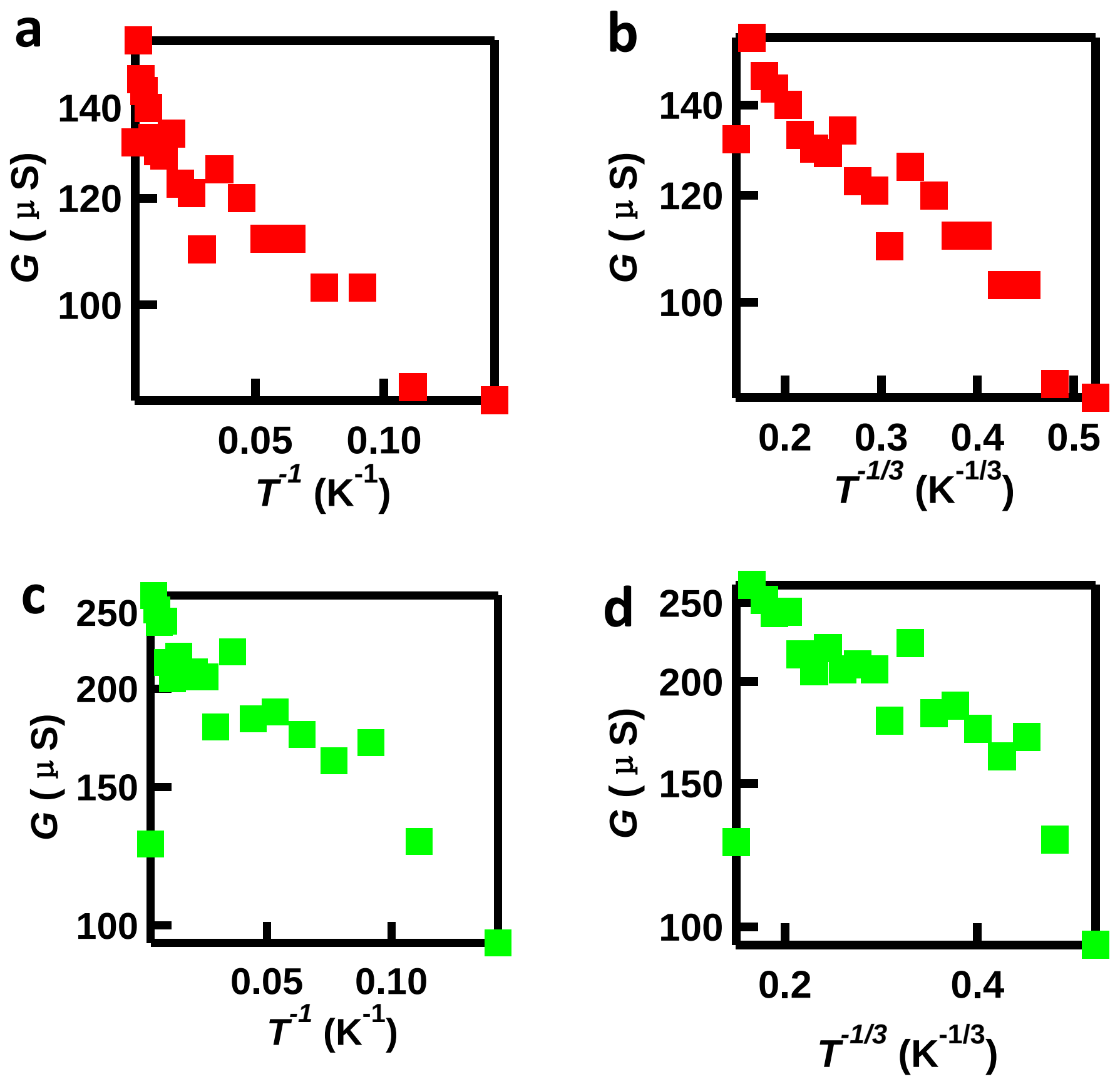

**Figure S2.** (a,b) Zero bias conductance, G at the Dirac point (at $V_g = 0$ V) vs $T^{-1}$ and $T^{-1/3}$ for a chromium (Cr) fuctionalized graphene device. (c, d) Zero bias conductance G at a highly doped regime (at $V_g = -42$ V) vs $T^{-1}$ and $T^{-1/3}$.

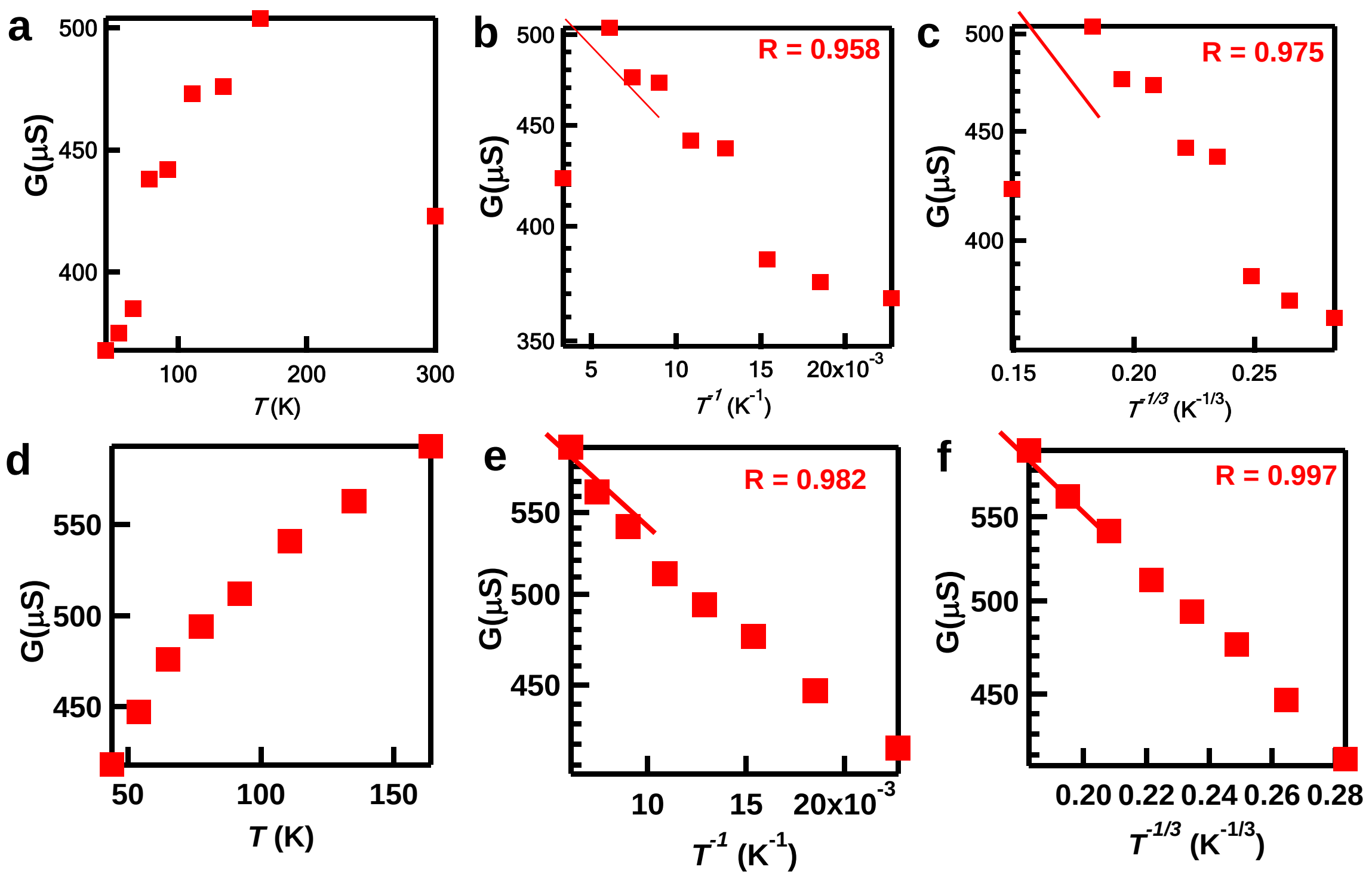

**Figure S3.** (a,b,c) Zero bias conductance, G at the Dirac point (at $V_g = 0$ V) vs T, $T^{-1}$ and $T^{-1/3}$ for another chromium (Cr) fuctionalized graphene device. (d,e,f) Zero bias conductance G at a highly doped regime (at $V_g = -56$ V) vs T, $T^{-1}$ and $T^{-1/3}$.

**XPS Charcaterization Data:** X-ray photoelectron spectroscopy (XPS) characterization was carried out by using a Kratos AXIS ULTRA$^{DLD}$ XPS system equipped with an Al K$\alpha$ monochromatic X-ray source and a 165-mm electron energy hemispherical analyzer. Vacuum pressure was kept below $3 \times 10^{-9}$ torr during the acquisition. The survey spectra were recorded using 270 Watts of X-ray power, 80 pass energy, and 0.5 eV step size. The high-resolution scans were run using power of 300 watts, 20 pass energy and step size of 0.05 eV. A low-energy electron flood from a filament was used for charge neutralization. The high-resolution scans were run using power of 300 watts, 20 pass energy and step size of 0.05 eV. The sensitivity factors used for the calculation of the elemental compositions are 0.278 for C1s and 2.427 for Cr2p.

As discussed in the main text of the manuscript because of the very small dimensions of the micromechanically exfoliated single-layer graphene (SLG) flakes and the presence of additional graphitic flakes on the silicon substrates is unavoidable, CVD-grown SLG (4 mm x 4 mm, on Cu-substrate) was prepared for the XPS experiments.

We analyzed two different functionalized samples, prepared by Methods B and C (see experimental details in the manuscript) and we obtained C : Cr ratios of 32:1and 18:1, respectively. **Figure S4** illustrates a survey spectrum and high-resolution spectrum of Cr2p.

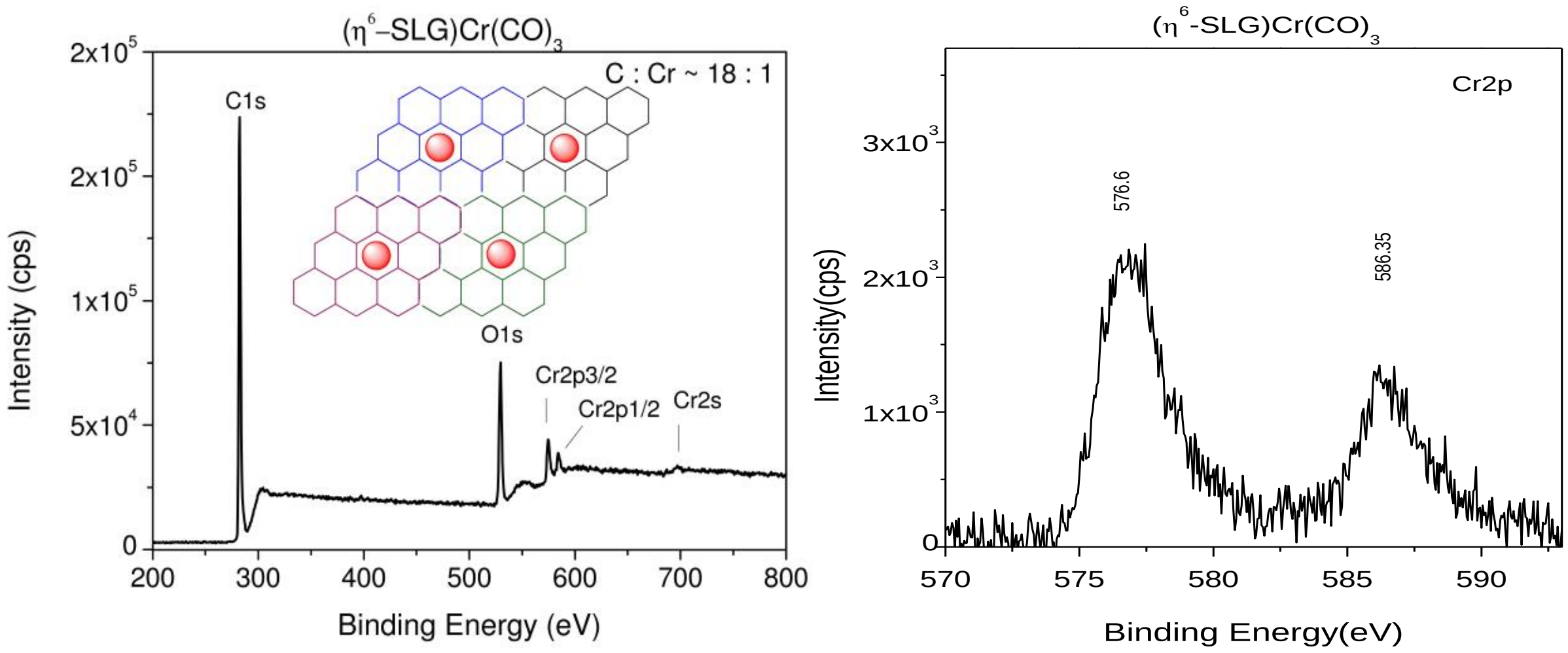

**Figure S4.** (Left) Survey spectrum of CVD-grown single-layer graphene (SLG) functionalized with chromium(0)tricarbonyl moieties. The inset shows the structure corresponding to the C:Cr ratio of 18:1 estimated from the C1s and Cr2p peaks, taking into account the sensitivity factors for carbon and chromium; the colored circles represent each of the hexahapto-bonded –$Cr(CO)_3$ moieties over the graphene surface. (Right) High resolution spectrum of Cr2p signals.